# Do we really need the anthropic principle?

V. Dorobanţu*

Politehnica University of Timişoara,
Physics Department, Timişoara, Romania,

**Abstract.**
*Dalai Lama, a very religious man, suggested at a conference of the Society for Neuroscience, held in Washington D.C., November 12, 2005, that we all should assume "a healthy dose of skepticism toward religious pronouncements". On the other hand, a scientist started calling the anthropic idea a principle. How can we deal with these statements?*

**Introduction**

As almost everyone knows, the cuckoo is a bird which lays eggs into other birds' nests. The host birds nurture the cuckoo's eggs as if they are their own. After hatching, the young cuckoos kick their nest mates out, remaining the only recipients of foster care. Where do we come from? Where are we headed? These are simple questions that lack simple answers. In the year 1973, at a conference held in Poland to celebrate Nicolaus Copernicus' $500^{th}$ birthday, Brandon Carter from Cambridge University placed the anthropic idea in the nest of science and elevated it to the status of a principle! What can the young cuckoo principle tell us?

Firstly, let's do a quick review of the anthropic idea.

1.a) Weak Anthropic Principle (WAP) [1] states that w*e must be prepared to take into account the fact that our location in the Universe is* necessarily *privileged to the extent of being compatible with our existence as observers.*

1.b) Strong Anthropic Principle (SAP) [1] states that *the Universe (and hence the fundamental parameters on which it depends) must be such as to admit the creation of observers within it at some stage.*

2.a) Barrow and Tipler [2] gave another formulation for WAP: *the observed values of all physical and cosmological quantities are not equally probable, but they take on values restricted by the requirement that there exist sites where carbon-based life can evolve, and by the requirements that the Universe be old enough for it to have already done so.*

2.b) For SAP, Barrow and Tipler [2] say: *the Universe must have those properties which allow life to develop within it at some stage in its history because:*
  i) *there is one possible Universe 'designed' with the goal of generating and sustaining 'observers';*
  ii) *observers are necessary to bring the Universe into being (Wheeler's Participatory Anthropic Principle (PAP));*
  iii) *an ensemble of other different universes is necessary for the existence of our Universe (which may be related to the Many_Worlds interpretation of Quantum Mechanics).*

*E-mail: vdorobantu@gmail.com

3.a) On the other side, regarding the WAP, Hawking [3] maintains that: *in a Universe that is large or infinite in space and/or time, the conditions necessary for the development of intelligent life will be met only in certain regions that are limited in space and time. The intelligent beings in these regions should therefore not be surprised if they observe that their locality in the universe satisfies the conditions that are necessary for their existence.*

3.c) For SAP Hawking [3] mentions: *there are either many different universes or many different regions of a single universe, each with its initial configuration and, perhaps, with its own set of laws of science. In most of these universes the conditions would not be right for the development of complicated organisms; only in the few universes that are like ours would intelligent beings develop and ask the question: "why is the universe the way we see it?" The answer is then simple: if it had been different, we would not be here!*

**Comments**

1 a) A formulation of WAP that includes language such as *we **must** be prepared, is reminiscent of* a warning. And, if we read further on that, *we are privileged* (and even *necessary*), it becomes an order: *you must be prepared,* otherwise *I shall cut the privilege*! A privilege is granted by someone. Guess by whom? Who can place us in the universe, by a *necessarily privilege*? Only God can have such power! The WAP is an appeal to deity.

1 b) The SAP is also imperative: *the universe **must** be such as to admit the creation of observers within it at some stage.* To observe what? The universe as a whole: do you see what I have done? Kurt Gödel [4], trying to prove the existence of God based upon Leibniz's concept of *positive* properties, defines*: something is God-like only if it posses all positive properties.* Is vanity, then, a positive property?

2a) *… the observed values of all physical and cosmological quantities are not equally probable..* There is no doubt: *…are not equally probable!* And the life, by all means, must be carbon – based. No other kind, only this!

2b) Barrow and Tipler are quite sure: *there is one possible Universe 'designed' with the goal of generating and sustaining 'observers'.* Yes! We, human beings, are the hub of the Universe. Even more, of all universes, because: *an ensemble of other different universes is necessary for the existence of our Universe.* What a huge expense !

3a), 3b) Hawking's tautological formulations reach the climax: the intelligent beings develop only to ask one question: *why is the universe the way we see it?* And an intelligent answer, suddenly, comes: *if it had been different, we would not be here!* No comment!

**About observers**

So far, no one has made a discovery that would suggest that Quantum Mechanics isn't correct. Bryce DeWitt [5], reasoning along these lines, found out that the wavefunction of the Universe is constant because the total Hamiltonian of the Universe is zero. Andrei Linde [6], trying to answer the question "why do we see the Universe evolving in time in a given way", says that we have to divide the Universe into two parts, one with an "observer" (or whatever it can be) having his own clock, and one that encompasses the rest of the Universe. Let's go a little bit further.

Let $|\Psi(\mathbf{x}, t)\rangle$ be the wavefunction of the Universe, depending on all **x** coordinates and time, and **H** the total Hamiltonian. Then, from the Schrödinger equation we have $\mathbf{H}|\Psi(\mathbf{x}, t)\rangle = 0$ because **H** = 0, which is the Wheeler - DeWitt equation. On the other side, we will get $|\Psi(\mathbf{x}, t)\rangle = \text{constant} = C_0$. Let's call the "initial" wavefunction of the Universe $|\Psi(\mathbf{x}, t)\rangle = \text{constant} = C_0$, the nothingness wavefunction.

Now, let's split up the total Hamiltonian $\mathbf{H} = \mathbf{H_o} + \mathbf{H_u}$ with $\mathbf{H_o}$ the Hamiltonian of the "observer", and $\mathbf{H_u}$ the Hamiltonian of the remaining Universe.

The "observer" and the remaining Universe will evolve so that the "observer" will be in the state $|\Psi_o(\mathbf{x_o}, t)\rangle$, and the Universe in the state $|\Psi_u(\mathbf{x_u}, t)\rangle$, and, as a consequence, the nothingness wavefunction will be the direct product of the two, $C_0 = |\Psi_o(\mathbf{x_o}, t)\rangle \otimes |\Psi_u(\mathbf{x_u}, t)\rangle$. With this assumption, the Schrödinger equation will split up in two equations: one describing the "observer's" evolution, and the other one the evolution of our universe. The "observer" is (at least at the beginning) completely independent, so, from the quantum point of view, $\mathbf{H_o}$ must be the Hamiltonian for a free particle, and, consequently, the solution of the corresponding Schrödinger equation is the wavefunction for a free particle. Now, from the nothingness wavefunction we can find out the wavefunction corresponding to the evolution of our universe.

The above dialectical exercise, or the itch to speculate, raises a question: is that **"observer",** from outside of the Universe, one with consciousness or not? Does it matter, or not? None of these possibilities can be excluded without narrowing our knowledge. Regardless whether it has or it doesn't have consciousness, the part that made up the "observer", has contributed to the birth of our universe.

Oa) If the observer was separated from our universe by his own will (therefore consciously), and tuned the initial conditions in order to see the Universe as we do today, then it seems that we are dealing with a God-like participant observer. However, we must mention that He took advantage of His power **only** in creating our universe. After the creation, He became an *external* observer. As an *external observer*, He can be:

Oa1) **completely free**, and in this case He **has no more interference** with our Universe and can no longer influence evolution. Otherwise, any subsequent **direct** intervention would have led to His reunification with our universe. The reunification would have triggered the remaking of the initial Universe, which involves the destruction of this Universe, and of Him, as a separated part, as well. Why? Because the nothingness wavefunction (3) will ask for the right to describe the situation and, as a consequence, the Universe will be a dead one. Why would one contribute to the creation of a universe, just to destroy it afterwards? Can it be a simple whim? This is not in accordance with the positive properties, and if at the beginning of this paragraph I was pretty close to giving some credit to the anthropic idea, now, because vanity (or whim) can hardly be considered a positive property, I have doubts;

Oa2) by His own consciousness He can **interfere with other universes**, which means that the Hamiltonian $H_o$ is changed, so the $H_u$, and consequently, the corresponding wavefunctions too, but in the frame of the initial nothingness wavefunction. This possibility, I mean Oa), may be a support to the anthropic principle, but from a physicist's point of view **there is no** "experimental" data encouraging such a speculation. **For a scientist**, the revelation **is too removed to be taken seriously**.

Ob) The second possibility consists in a fortuitous separation of a part from the initial universe, and this part may be a "particle" with its own lifetime (and clock). I think that such a situation is more physical, and in accordance with the theories of chaotic inflation [6]. Furthermore, that particle separated from our universe can move freely or, by chance, to meet another universe, in which case the above discussion **Oa1) and Oa2) hold true, without involving any consciousness,** and obviously, **no deity is necessary**.

According to **Oa2), with no deity,** the subsequent meetings lead to a perpetual creation of universes, which is thus, regulated by chance, and as a consequence, so are the corresponding changes in our own universe. We, human beings, may be considered as observers, **inside** observers, and even participant observers as Wheeler said, **but not observers as a solely final goal**. When we say final, one means finish, ready, you have no chance, that's all! Have we been created to reach such a conclusion only? Pathetic!

Again, the anthropic idea is not going to convince.

**Conclusions**

The anthropic idea, *per se,* is simply an idea, but it fuelled all sorts of speculation simply because it originated within the realm of science. Herein lies the problem: religions, regardless of their nature, speculate as much as possible on the basis of ideas that come from the scientific world. They even go as far as giving a "scientific" reasoning for their actions. Some do it shamelessly: on American campuses, an organization named IDEA (Intelligent Design and Evolution Awarness) is gaining in popularity. The leader of this organization, taking advantage of the fact that he is the holder of three degrees (one of them being in Mathematics), explained that he didn't find a viable explanation of life's appearance on Earth without introducing the idea of an intelligent designer. A stance such as this one can be convincing when the holder of three university degrees is addressing an audience of 18 or 19-year olds, who are still forming their opinions. Introducing religious elements in education is damaging. Furthermore, the Kansas State Board of Education [8] recently voted to allow intelligent design to be taught in public schools. That's why Eugenie Scott, the executive director of the National Center for Science Education, is to be appreciated in her efforts to fight science's opponents.

The main danger of ideas like the anthropic one or intelligent design, which, in fact, are two names for the same thing, consist of distorting young people's minds in order to estrange them from science and, by consequence, bring them closer to religion. Religion is a very personal matter, and schools (be they public or private) must not be involved in students' personal beliefs. We know very well that when religion turns into politics, then, crime is very close. Two examples:

C1) During the Middle Ages, the Inquisition made thousands of crimes in the name of God. **Giordano Bruno**, for instance, *Filosofo, arso vivo a Roma, per volonta del Papa, il Febbraio 1600,* **because there was no room in his cosmology for the Christian notion of divine creation.** Also, Galileo Galilei, defending the Copernican planetary system, was condemned to lifelong imprisonment by the Inquisition. The Pope John Paul II, in the name of the Catholic Church, asked forgiveness, but almost 400 years later.

C2) Nowadays, Islamic fundamentalism encourages crimes in the name of God, going as far as to recruit suicide bombers by telling them they will have a beautiful life in Heaven ("In them will be Companions, good, beautiful, whom no man or Jinn before them has touched", Koran 55.70, 55.74)!?

I can also recall the reckless hate provoked by some cartoons.

Perhaps in another 400 years an Islamic equivalent of the Pope John Paul II will ask forgiveness. Won't it be too late? Just like that: aren't we able to learn from History? That can happen only if religion is mixed with the lack of scientific education and takes advantage of that lack. **The education must be secular**.

To answer the question raised in the title of this article: the anthropic idea proved to be more useful in speculating outside of science than in science, as H. Ross [9] does, or in " Anthropic Philosophy" as Wm. L. Craig countenances [10]. As we already know, the evolution of our universe from the Big Bang up to now doesn't need any divine intervention. To speculate that the external observer is of a divine origin is not sustained by any data. I still think, therefore, that the anthropic idea is the cuckoo bird's egg in science's nest.

I believe that it is necessary to remember the Dalai Lama's urge regarding *a healthy dose of skepticism toward religious pronouncements.*